\documentclass{raa}            

\usepackage{graphicx,times}             
\usepackage{natbib}
\usepackage{amssymb,amsmath}
\bibpunct{(}{)}{;}{a}{}{,}

\begin{document}

   \title{Disk evolution of the M87's nucleus observed in 2008
}

   \volnopage{Vol.0 (20xx) No.0, 000--000}      
   \setcounter{page}{1}          

   \author{Fei Xiang
      \inst{1}
   \and Cheng Cheng
      \inst{2}
   }

   \institute{Department of Astronomy, Yunnan University, Kunming 650091, China, {\it xiangf@ynu.edu.cn}\\
        \and
            Chinese Academy of Sciences South America Center for Astronomy, National Astronomical Observatories, CAS, Beijing 100101, China, {\it chengcheng@bao.ac.cn}\\
 \vs\no
   {\small Received~~20xx month day; accepted~~20xx~~month day}}

\abstract{We report the discovery of year-scale X-ray variation in
the nuclear region of the M87 by reanalyze the 8 Chandra
observations from 2007 to 2008. The X-ray spectra are fitted and
decomposed into disk and flaring components. This year-scale X-ray
variability can be explained quite well by a simple clumpy
accretion model. We conclude that the central super-massive
blackhole of the M87 was accreting a cloud of $\sim0.5$M$_\odot$
at that time.
 }

   \authorrunning{Fei Xiang \& Cheng Cheng }            
   \titlerunning{accretion disk evolution in M87}  

   \maketitle

%
%
\section{Introduction}           
\label{sect:intro} The nucleus of M87 galaxy is an ideal object
for studying gas accretion into central Super-Massive Black
Hole(SMBH). The proximity ($\sim$ 16 Mpc) of M87 makes its central
tens-parsec region resolvable in various wavelength (e.g., Optic:
Perlman et al. 1999, radio: Biretta et al. 1995, x-ray: Marshall
et al. 2002). Based on the $Chandra$ observations, Di Matteo et
al. (2003) obtained that the mass accretion rate
($\dot{\textrm{m}}$) is $\sim10^{-3}$ of Eddington accretion rate
($\dot{\textrm{m}}_{\tiny\textrm{Edd}}$) and the radiation
efficiency ($\eta$) is $\sim10^{-5}$. Recently, Levinson \& Rieger
(2011) suggested that the ratio of $\dot{\textrm{m}}$ over
$\dot{\textrm{m}}_{\tiny\textrm{Edd}}$ should be as low as
$\sim10^{-4}$. These observations implied that the central region
of M87 is under the Radiation Inefficiency Accretion Flow (RIAF)
mode.

The Advection Dominated Accretion Flow (ADAF, Narayan \& Yi 1994)
model, the most widely applied RIAF model, has successfully
explained the emission from Low Luminosity AGNs (LLAGNs) and been
applied to M87 (Ho 2008, Nemmen et al. 2014, see Yuan \& Narayan
2014 for a detail review). By analysis of the Spectral Energy
Distribution (SED), it was found that the X-ray emission from
M87's nucleus could be mainly attributed to ADAF component (Di
Matteo et al. 2003, Wang et al. 2008, Li et al. 2009, Cui et al.
2012, Nemmen et al. 2014).

For the LLAGNs, one of the interesting issues is the evolution of
the accretion mode. Theoretical models have proposed that the gas
flow in LLAGNs might be clumpy, which could lead to some
year-scale variation in mass accretion rate (Yuan 2003, Wang et
al. 2012). The variation of mass accretion rate could be detected
through the variation of X-ray luminosity, provided the bolometric
luminosity of pure ADAF can be estimated by X-ray luminosity
multiplying a bolometric correction (Elvis et al. 1994, Ho 1999,
Hopkins et al. 2007). The year-scale variation of X-ray flux has
been found in M81 with Swift/XRT (Pian et al. 2010). Considering
the similarity of LLAGNs and the larger scale of the M87's
accretion flow, it can be expected that such variation might be
found in M87, which might give some constraints on inhomogeneous
accretion model.

$Chandra$ observations have revealed month-scale variations of the
nuclear X-ray luminosity in M87 (Harris et al. 2009), which may be
accompanied by the change of the X-ray spectrum (Hilburn \& Liang
2012). Although the X-ray emission in the `low state' (i.e. low
luminosity) of M87 can be explained by pure ADAF model (Di Matteo
et al. 2003, Wang et al. 2008, Li et al. 2009). The X-ray emission
in `high state' (i.e. high luminosity) might have more complex
components and behaviors due to the occasion of the mini-jet
(Giannios et al. 2010, Cui et al. 2012). Thus, it is necessary to
check the X-ray contents of `high state' carefully.

 In addition, M87 is the unique non-blazar radio galaxy detected to
emit Very High Energy (VHE) gamma rays. In Feb 2008, a giant VHE
flare was observed in M87 by MAGIC and VERITAS respectively
(Albert et al. 2008, Acciari et al. 2009). Subsequently, a
$Chandra$ observation to M87's nucleus was performed 3 days after
the VHE flare and a sharp enhancement in the nuclear X-ray
luminosity was observed, indicating that the location of the VHE
flare was in the M87's nucleus (Harris et al. 2009). In 2010,
another VHE flare was captured. With the help of subsequent
multi-wavelength observations (VHE: VERITAS, MAGIC, H.E.S.S;
X-ray: Chandra; radio: 43GHz, VLBA; Aliu et al. 2012, Abramowski
et al. 2012, Hada et al. 2012), the location of the flare was
confirmed to be near the blackhole, which suggests that the VHE
and X-ray flares come from the outburst near the SMBH (Cui et al.
2012). This succeeding variability in $\gamma$-ray and x-ray bands
might offer us more hints about the evolution of nucleus in M87.

In this paper, We re-analyze a series of $Chandra$ observations
from July 2007 to August 2008 (PI: Dr. John Biretta) and find a
year-scale variability component which could be explained by a
simple inhomogeneous accretion model. This paper is arranged as
follows, data analysis is described in section 2, the spectra
fitting results are presented in section 3, the clumpy accretion
and the contents of X-ray emission are discussed in section 4 and
finally the conclusions come in Section 5.

\section{Data analysis}
\label{sect:Data} From 2007 to 2008, eight $Chandra$ observations
to M87 (Observation ID: from 7354 to 8581) were proposed by Dr.
John Biretta, aimed to monitor the HST-1 and revealed nuclear
X-ray brightening after the VHE flare (Harris et al. 2009). The
HST-1 is a knot in M87'jet. It was revealed by the HST (Boksenberg
et al.1992) and draw the attention from x-ray to radio band soon
after the discovery. The HST-1 is much brighter than the nucleus
and shows strong variability in X-ray. Its X-ray luminosity had
been increasing from 2001 to 2005 and started to decrease after
2005. Its peak luminosity in 2005 could reach about 10 times of
the nuclear luminosity. Because the location of HST-1 is near the
nucleus (65 pc projected), it may produce the `light pollution' on
nucleus (Harris et al. 2003, 2006, 2009). To avoid strong `light
pollution', we choose the data from 2007 to 2008 for analysis. At
that time, the HST-1's x-ray luminosity became comparable to that
of nucleus. The possible `light pollution' from HST-1 will be
discussed in later paragraphe . The livetime for each observation
is about 4.7 ks, the observation instrument is ACIS-I, and the
observation mode is FAINT. Under the FANIT mode, the frametime is
$0.4$ s to avoid the significant pileup effect. Even for the
brightest observation (ID: 8577), the peak counts rate of the
nucleus is $\sim0.2$ counts s$^{-1}$pixel$^{-1}$. The counts rate
per frame is $\sim0.08$ counts s$^{-1}$. At this level, the pileup
fraction is less than $5\%$ (Davis 2001). The peak counts rates of
the other observations are around $0.1$ counts
s$^{-1}$pixel$^{-1}$. So the following spectral fittings is free
of the pileup effect.

In our analysis, The level 2 data is processed with CIAO 4.2 and
CALDB 4.2.0, which is produced through the Standard Data
Processing(SDP), in which good time intervals (GTIs) filtering,
cosmic ray rejection, and position transformation are performed.
The structure of nucleus and jet is clear in the images of these
observations (e.g. ID: 8576 at Jan. 4th 2008 in Fig 1). The center
of the X-ray nucleus is located at
R.A.$=12^{\tiny\textrm{h}}$30$^{\tiny\textrm{m}}$49$^{\tiny\textrm{s}}.42$,
Dec.$=12^{\circ}23'28''.05$ (J2000) which is consistent with that
of previous observations (Wilson \& Yang 2002). To extract the
spectra of nucleus, we select the box region including nucleus
($1''.8\times2''.3$) as the source region for every observation
(Fig 1). The location of source regions are chosen on the basis of
contours (e.g. bottom panel in Fig 1). To eliminate the
contribution of foreground and background, we select a box region
beside the source ($17''.0\times8''.5$) as the background region
(top panel in Fig 1), where no clear point source has been
detected. The spectra, Ancillary Response Files (ARFs) and
Response Matrix Files (RMFs) are generated through the
`specextract' script in CIAO. The size of source region is
comparable to the size of Point-Spread Function (PSF). The
`arfcorr', which is performed automatically in `specextract'
script, will give the correction factors on ARFs. In corrected
ARFs, the column `PSF FRACTION' presents the fraction of PSF
counts within the selected region at each energy band
(http://cxc.harvard.edu/ciao/threads/pointlike/). We checked the
PSF Fraction in every corrected ARFs, ensuring the PSF Fraction is
from 0.85 (hard band) to 0.95 (soft band). The contribution of
adjacent HST-1 has been estimated to be about $6\%$ by Harris et
al. (2006).

\begin{figure}
\centering
{\includegraphics[scale=0.458]{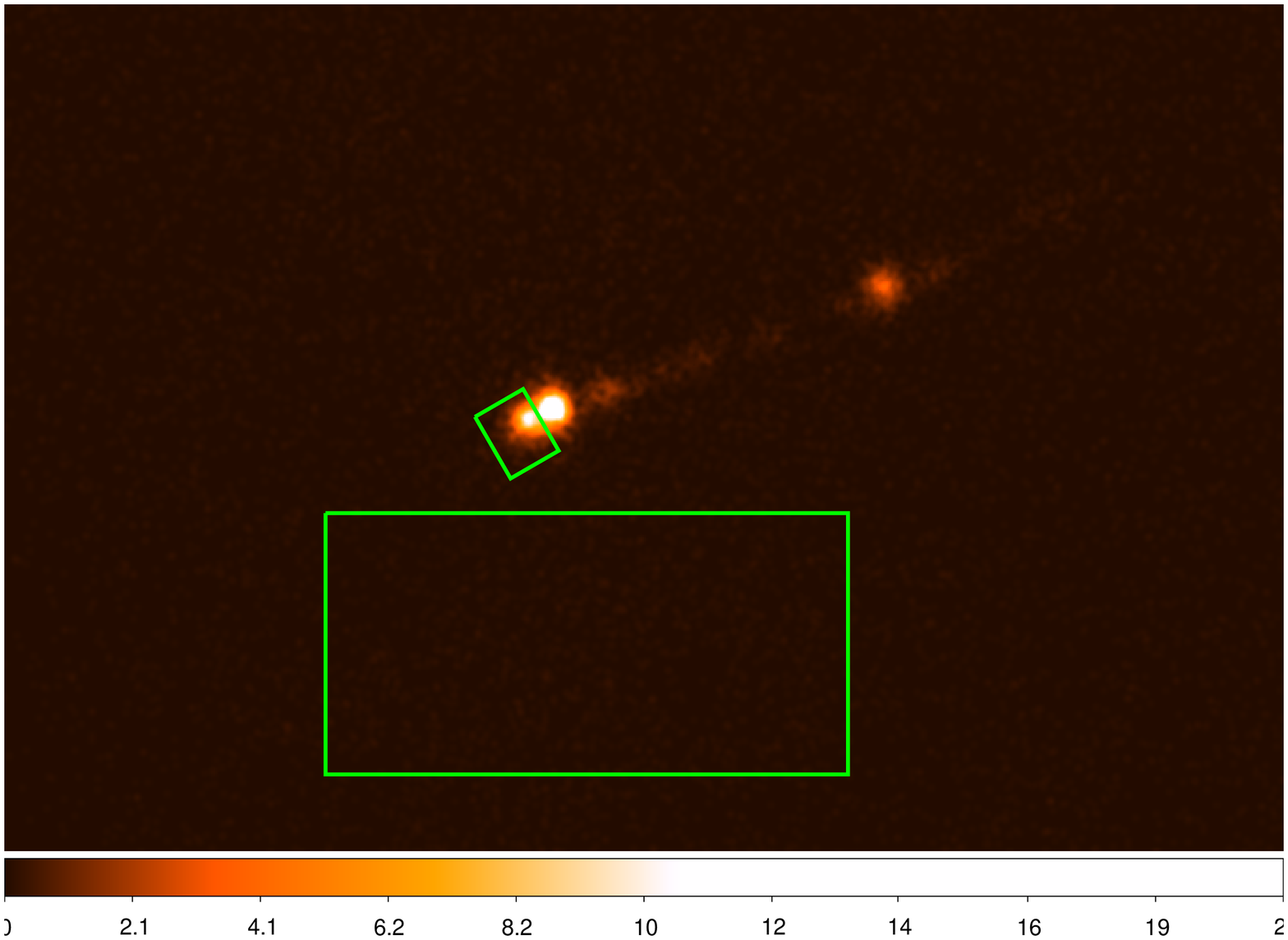}}\\
\includegraphics[scale=0.20]{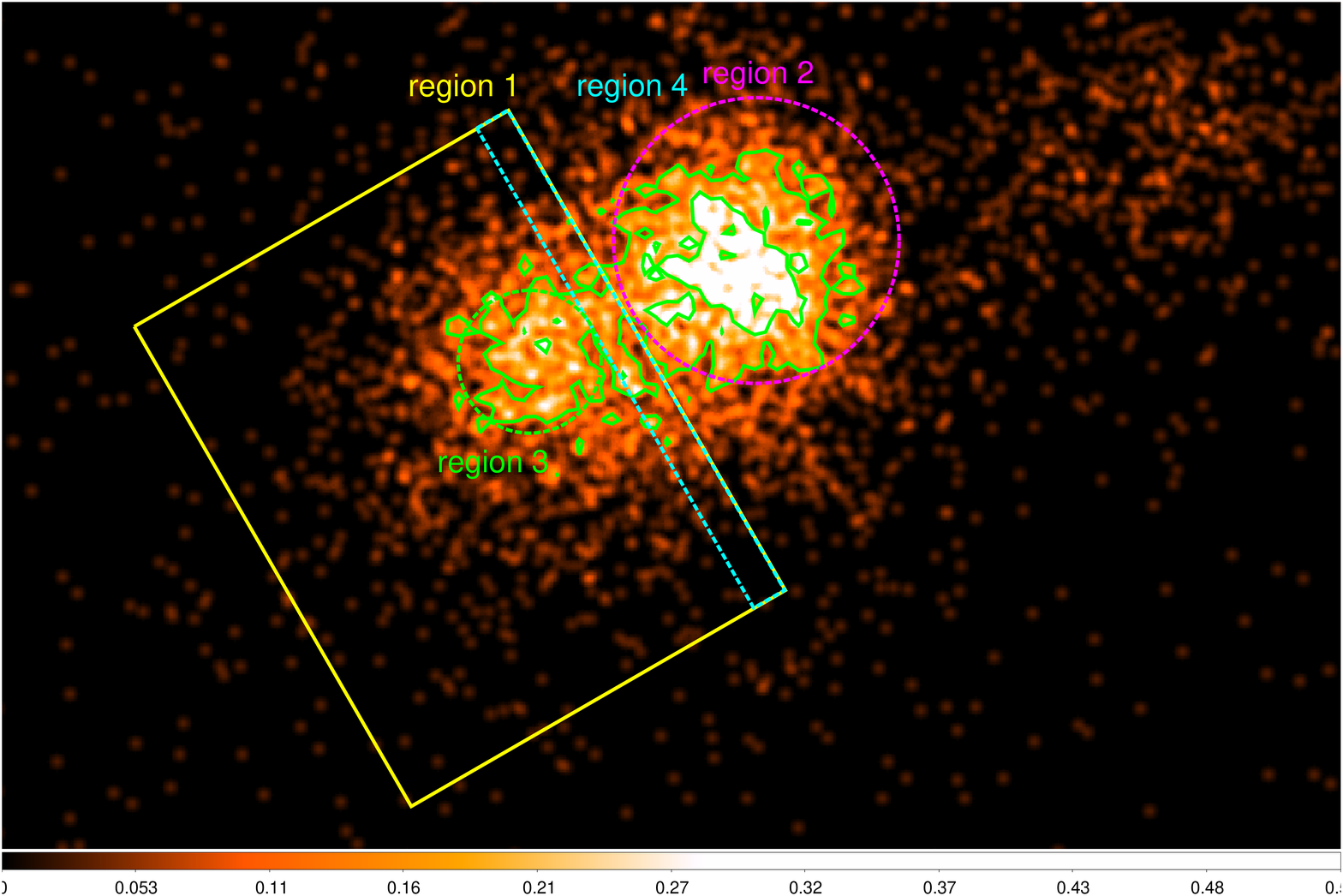}
\caption{The X-ray image of M87's jet observed on Jan 4th 2008. In
top panel, the small box indicate the source region and the big
box is the background region; the left point source is the nucleus
and the right point source is the HST-1. Bottom panel shows that
the box of the source region is chosen on the basis of contours.
The contour levels are 0.12, 0.24 and 0.48 counts per bin. Bin
size is pixel/64 and a Gaussian smoothing with kernel radius of 5
bins is applied. In bottom panel, yellow solid box: source region
(region 1); magenta dotted circle: HST-1 (region 2); green dotted
circle: nucleus' center (region 3); cyan dotted strip: region
between nucleus' center and HST-1 (region 4).}\label{fig1}
\end{figure}

\section{Results}
\label{sect:re} Previous observations to M87's nucleus indicate
that the nuclear X-ray spectra could be fitted well with an
absorption-modified powerlaw model (Wilson \& Yang 2002). So we
use XSPEC (version 12.6.0) to fit the spectra with the model as
follows (Arnaud 1996):
\begin{equation}\label{eq1}
Model_1=Wabs(n_\textrm{H})\times Powerlaw(\Gamma,\textrm{norm})
\end{equation}
The equivalent hydrogen column density ($n_{\tiny\textrm{H}}$) is
fixed as $\sim6.1\times10^{20}$cm$^{-2}$ in the spectra fitting of
this work, which is derived by Wilson \& Yang (2002). The
resulting photon index, normalization of the spectra and flux
(2-10keV) are given in Table 1. The spectral variations are shown
in Fig 2, which reveals a month-scale variability. The nuclear
luminosity increases sharply and then returns to the original
level in a scale of month with the photon index changing in
reverse direction.
\begin{table}
\begin{center}
\caption{The parameters of spectra fitting with absorbed power
law. The error bars are calculated with the confidence of $68\%$ }
\label{p12}
\begin{tabular}{@{}llllll}
\hline \hline

 ID& Photon index & norm &flux$_{\tiny\textrm{2-10keV}}$ &$\chi^2$/DOF ~~~\\
               &              & ($10^{-5}$)&($10^{-12}$erg/s/cm$^2$)     &\\
\hline

7354 &$2.29\pm0.07$  & $39\pm2$  & $0.66\pm0.06$ &12.58/27\\
8575 &$2.02\pm0.05$  & $69\pm2$  & $1.70\pm0.10$ &40.53/37\\
8576 &$2.10\pm0.05$  & $74\pm2$  & $1.62\pm0.12$ &37.69/40\\
8577 &$1.71\pm0.03$  & $119\pm3$ & $4.74\pm0.20$ &63.83/65\\
8578 &$1.78\pm0.04$  & $74\pm2$  & $2.68\pm0.14$ &49.29/46\\
8579 &$2.12\pm0.05$  & $69\pm2$  & $1.47\pm0.10$ &36.56/37\\
8580 &$1.75\pm0.04$  & $91\pm2$  & $3.38\pm0.15$ &59.13/55\\
8581 &$2.15\pm0.05$  & $57\pm2$  & $1.16\pm0.08$ &69.01/74\\
\hline
\end{tabular}
\end{center}
\medskip
\end{table}

\begin{figure*}
\includegraphics[scale = 0.95]{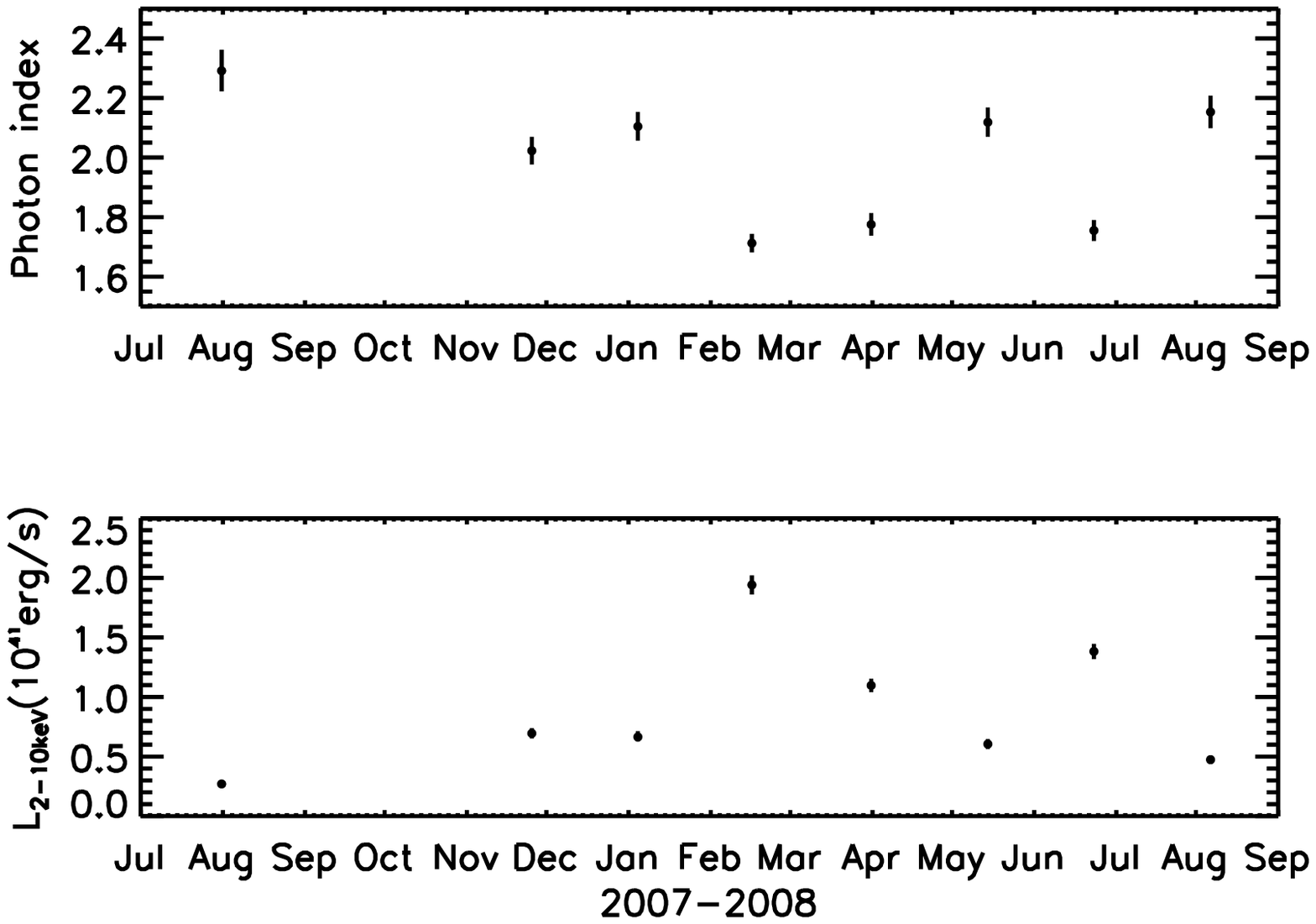}
 \caption{The variation of the X-ray spectra parameters observed in M87's nucleus
 from July 2007 to September 2008.}\label{fig2}
\end{figure*}
The 2-10keV luminosity light curve in Fig 2 is consistent with the
photometric light curve obtained by Harris et al. (2009). The Fig
3 illustrates the relation between the photon index ($\Gamma$) and
flux, from which we could divide the states of nucleus into two
types. The five observations with the photon index about 2.1 and
low flux are denoted as the `first class', which are shown by
triangles in Fig 3; the rest three observations with the photon
index around 1.75 and relatively high flux are denoted as the
`second class' (dots in Fig 3).

\begin{figure}
\center{\includegraphics[scale=0.6]{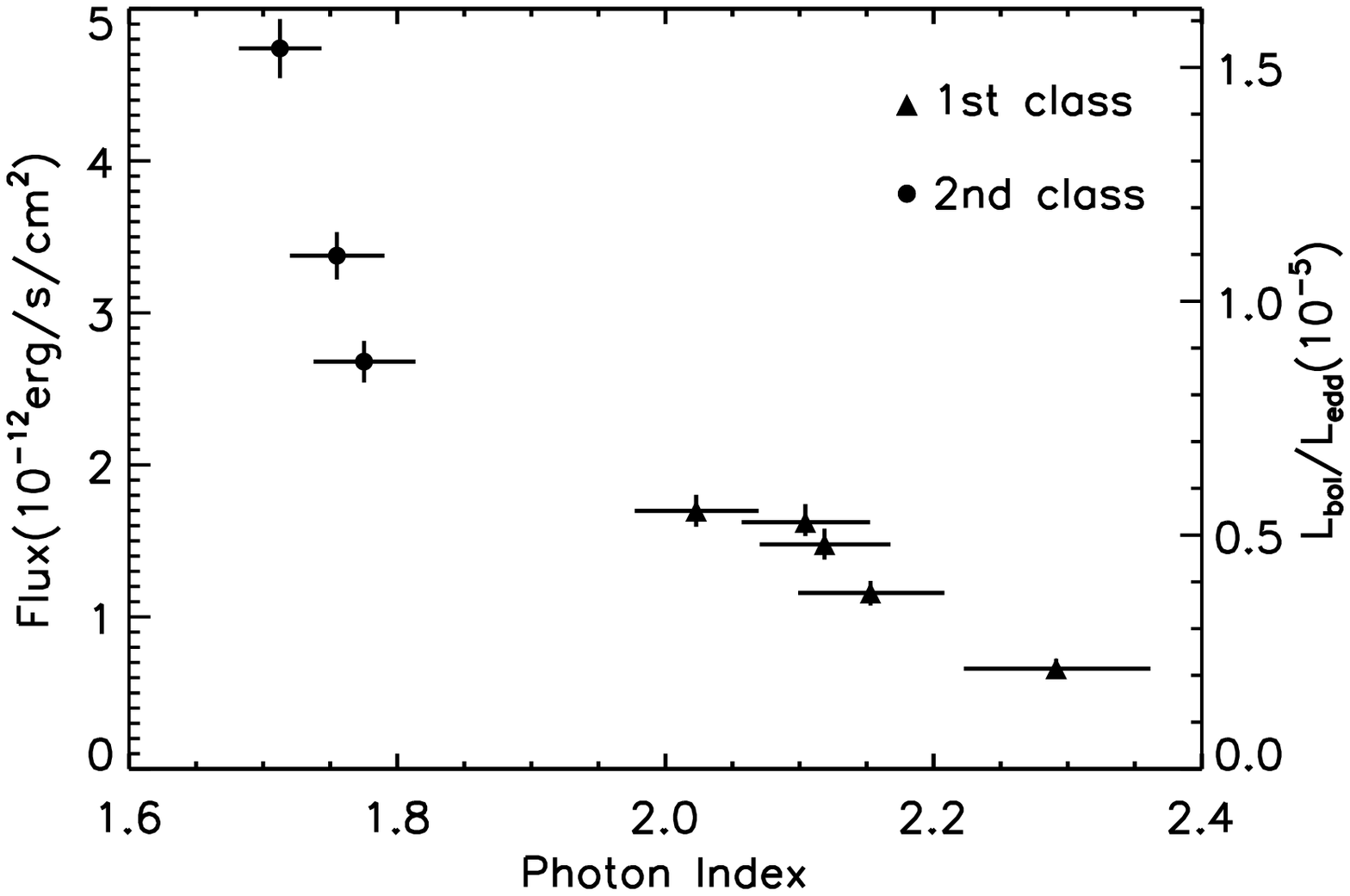}}
 \caption{The relation between the X-ray photon index and the flux. The $L_{\tiny\textrm{Bol}}$ is calculated by
$L_{\tiny\textrm{{Bol}}}/L_{\tiny\textrm{2-10keV}}\sim30$ (Elvis
et al. 1994, Hopkins et al. 2007) and the $L_{\tiny\textrm{Edd}}$
is calculated with the black hole mass of $3\times10^9$M$_\odot$
(Fabian \& Rees 1995, Macchetto et al.1997). }\label{fig3}
\end{figure}

For the first class, the previous studies have confirmed that
their SED could be explained by pure ADAF model (Di Matteo et al.
2003, Li et al. 2009, Cui et al. 2012). And the latest SED fitting
by the model of ADAF plus jet also suggests that ADAF component
dominates the X-ray emission (Nemmen et al. 2014). In addition,
recent studies of polarization show that the optical polarization
of the nucleus is much lower than that of HST-1 in M87, which
implies that the radiative mechanism of the nucleus may be
different from the HST-1's (Perlman et al. 2011, Adams et al.
2012). Based on these results, it is likely that the X-ray
emission of the first class is mainly attributed to the ADAF
component. For the second class, the X-ray spectra become harder
with the increasing of flux. In the observation of 02/16/2008,
X-ray luminosity brightened up after the VHE flare. So it is
plausible that the flares occurred in these observations. We could
induce that there are two components in their X-ray emission: one
is the flaring component and the other is the quiescent component
whose spectrum is similar to that of the first class.

The variations of the X-ray emission in M87 nucleus also can be
divided into two types (Fig 2, Fig 5). Firstly, the flares occur
in the scale of month and are corresponding to the spectra
changing from the first class to the second class. Secondly, the
nuclear luminosity of the first class exhibits a year-scale
evolution (triangles in Fig 5), which shows the trend of fast
rising ($\sim4.2\times10^{40}$ erg/s, $10\sigma$) in 4 months and
then slow fading ($\sim2.2\times10^{40}$ erg/s, $5\sigma$) in
about 8 months. We have checked the variability of the first class
in Fig 5 through reduced Chi-square test. For all five points, the
probability (P value) of no variation is less than $10^{-5}$. So
it could be affirmed that the year-scale evolution of X-ray
emission exists in M87. Considering the complexity of the second
class, we only consider the results of the first class when
investigating the evolution of ADAF content(triangles in Fig 5).

In our analysis, the $n_{\tiny\textrm{H}}$ is fixed as
$6.1\times10^{20}$ cm$^{-2}$. The range of $n_{\tiny\textrm{H}}$
given by Wilson \& Yang (2002) is $4.7\sim7.6\times10^{20}$
cm$^{-2}$ ($90\%$ confidence), showing an excess above the
Galactic value ($2.5\times10^{20}$ cm$^{-2}$, Stark et al. 1992).
But Harris et al. (2006) demonstrate that the derived
$n_{\tiny\textrm{H}}$ highly depends on the fitting model. To
demonstrate the impact of the uncertainty in
$n_{\tiny\textrm{H}}$, we change the $n_{\tiny\textrm{H}}$ in
fitting from $2.5\times10^{20}$ to $8.5\times10^{20}$ cm$^{-2}$.
The reduced $\chi^2$ of these fittings are still acceptable
($\leq1.1$). The variation of $n_{\tiny\textrm{H}}$ could cause a
change of $\pm6\%$ on the derived flux and $\pm0.1$ on the photon
index. The uncertainty in $n_{\tiny\textrm{H}}$ may blur the
difference of photon index between the first and the second class.
But the variation of $n_{\tiny\textrm{H}}$ can only produce the
variation of $\pm6\%$ on luminosity and can not explain the
luminosity evolution observed in the first class. Provided that we
still use the same $n_{\tiny\textrm{H}}$ in 8 spectra fittings,
our conclusion will not be apparently affected.

Another factor to be taken into account is the `light pollution'
of HST-1. Firstly, the source regions in data analysis are chosen
on the basis of contours. In the 5 observations when the nucleus
is bright (ObsID: 8575,8576, 8577, 8578, 8580), the separation of
contours is clear but there still is about $6\%$ of the light
counts that come from HST-1 (Harris et al. 2006). For the 3
observations in which the emission of nucleus is weak (ObsID:7354,
8579,8581), the boundary is blurry so that the source regions may
include larger fraction of counts from HST-1 than the 5 other
observations. In this case, the actual nuclear emission in these 3
observations may be weaker and the year-scale variation may be
more intense than what has been observed. Secondly, the detections
of nucleus and HST-1 may suffer from a mutual effect of pileup
called ``Eat Thy Neighbor'' (Harris et al. 2006, 2009). When two
photons from each other come within $3\times3$ pixel grid at the
same frame, they will be considered as one event and the event
position is determined by the harder photon. If HST-1 become
brighter, the nuclear emission will be eaten more. As result, we
can only choose the observations in which HST-1 is not so bright
for analysis. Finally, if pileup is significant, it will distort
the PSF, which is called ``second-order effect of pileup'' in Harris
et al. (2009). The second order effect is hard to be estimated. A
possible way of reducing this effect is to select the observations
in which the pileup effect is not too strong for nucleus and
HST-1.

Because the complexity for estimating the contamination directly,
to determine the influence of HST-1 to our results, we check the
correlation of light curves between different regions with the
HST-1. We choose four regions to derive the counts rate light
curve. The first region is rectangle source region (region 1,
$1''.8\times2''.3$); the second region is the circle region at
HST-1 (region 2, radius: 0.6"); the third region is the circle
region at nucleus' center (region 3, radius: 0.3"); the forth
region is the strip region between HST-1 and nucleus' center
(region 4,$0''.15\times2''.3$ ) (Fig 1). The light curves of the
four regions are shown in Fig 4. We found that the light curve of
region 2 (HST-1) has an opposite trend with those of region 1
(source region) and region 3 (central nucleus). Their correlation
coefficient is -0.56, which is similar with the Harris et al.
2009's result. The light curves of region 1 and region 3 have the
same trend and the correlation coefficient is 0.99. The region 4
(strip region) is the region which may subject to the 'light
pollution' most possibly. But in Fig 4, its counts rate changes
slightly. Its correlation coefficient with HST-1 is -0.45 and that
with the nucleus is 0.81. The above correlation coefficients
suggest that change of the source region is determined by the
nucleus and the strip region is influenced by HST-1 little.

\begin{figure}
\center{\includegraphics[scale=0.6]{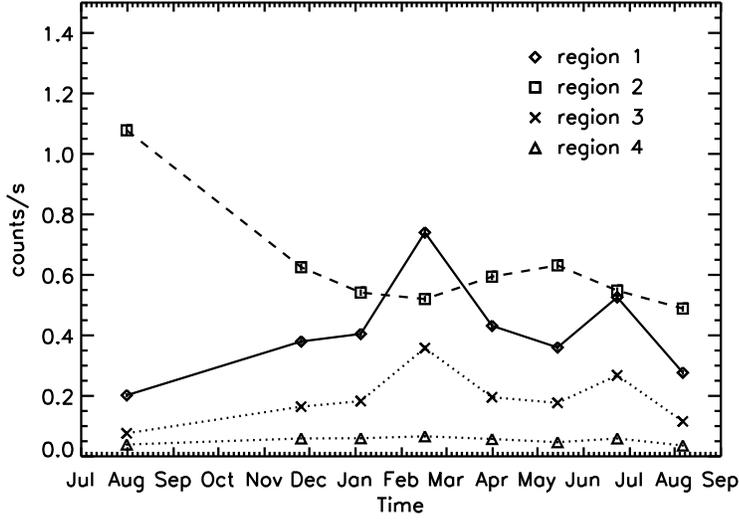}}
 \caption{The light curves of the above four regions.(Diamond: region 1;
 Square: region 2; Cross: region 3; Triangle: region 4.)  }\label{fig4}
\end{figure}

Further, if we assume that the counts rate of strip region is
influenced by HST-1 and the nucleus simultaneously, the counts
rate of the region 4 could be expressed as $Z=A*X+B*Y+C$: the $X$,
$Y$, $Z$ are the counts rate of region 2, 3, 4 respectively and
$A$, $B$, $C$ are the constants. We could conduct a multivariate
fitting to above light curves. We derive $A$, $B$ and $C$'s
values: 0.0002, 0.1 and 0.033. The very low value of coefficient
$A$ indicates little influence from HST-1, which confirms the
above conclusion.

\section{Discussion}
\label{sect:discussion}
\subsection{Clumpy accretion in M87}
For the first class (quiescent state), the nuclear luminosity
shows a slow evolution within a year (triangles in Fig 5). For all
five points, the disk luminosity was the lowest in the observation
made on 07/31/2007 and then increased quickly to the highest in
early 2008, reaching $\sim7.0\times10^{40}$ erg/s (the increment
$\sim10\sigma$). After that, in observation of late 2008, the
nuclear luminosity decreased to $\sim4.7\times10^{40}$ erg/s
slowly (the decrement $\sim5\sigma$). We define the mass accretion
rate as $\dot{\textrm{m}} = L_{\tiny\textrm{x}}/(\eta c^2) $,
where $\eta$ is the X-ray radiation efficiency. Di Matteo et al.
(2003) estimate that the
 average mass accretion rate of M87 is $\sim0.1$ M$_\odot/$yr and $\eta$ is
 about $10^{-5}$ through detailed study on the hot
 interstellar medium in M87. Based on these results, the above variation of
 X-ray luminosity could be attributed to the variation of mass
 accretion rate ($\dot{\textrm{m}}$).

The inhomogeneous accretion flow, which is called clumpy
accretion, might account for such variability. The gas clumps
(clouds), originated from gravitational or thermal instability in
accretion flow, might lead to a modulation in luminosity
(Ishibashi \& Courvoisier 2009, Strubbe \& Quataert 2009, Wang et
al. 2012). When a clump falls toward the central black hole, it
will be disturbed by the tidal force and form a gas ring of high
density. Basic equation for such an accretion ring is
\begin{equation}\label{sigma_eq}
    \frac{\partial \Sigma }{\partial t} = \frac{3}{R}\frac{\partial  }{\partial R}
    \left( R^{1/2}\frac{\partial \nu\Sigma R^{1/2} }{\partial R} \right),
\end{equation}
with initial matter distribution is:
\begin{equation}
    \Sigma(x, \tau= 0) =\Sigma_0\delta\left( R-R_{0} \right) ,
\end{equation}
here $\Sigma$ is the surface density of the accretion flow, $R$ is
the radius, and $\nu$ is kinematic viscosity parameter (Lin \&
Pringle 1987). $\Sigma_0$ is the initial surface density of the
accretion ring and the $R_0$ is the radius where the ring forms.
The $x=R/R_0$ presents the dimensionless distance to central black
hole and $\tau =(t-t_0)/\tau_0$ where the $t_0$ is the date when
the clumpy accretion started and $\tau_0$ is time scale of gas
falling. For simplicity, we take $\nu$ as a constant, and thus
analytical solution to Equation 5 can be written as(Frank et al.
2002):
\begin{equation}
    \Sigma(x,\tau) = \sigma_0\frac{1}{\tau}\frac{1}{x^{1/4}}\bm{e}^{-\frac{1+x^{2}}
    {\tau}}I_{1/4}\left( \frac{2x}{\tau} \right) ,
\end{equation}
and
\begin{equation}
    \tau_0=\frac{R_0^2}{12\nu}
\end{equation}
 Here $I_{1/4}$ is the modified Bessel function. The falling
velocity $v_{_R}$ could be derived through:
\begin{equation}
v_{_R} =\frac{3}{\Sigma R^{1/2}}\frac{\partial}{\partial R}(\nu
\Sigma R^{1/2}).
\end{equation}
So
\begin{equation}
    v_{_R} = - \frac{3\nu}{R_{0}} \frac{\partial  }{\partial R}
    \left[ \frac{1}{4} - \frac{1+x^{2}}{\tau} + \ln I_{1/4}\left( \frac{2x}{\tau} \right)
    \right],
\end{equation}
asymptotically, the radial velocity is
\begin{equation}
    v_{_R} \simeq -\frac{3\nu}{R_{0}}\left( \frac{1}{2x}-\frac{2x}{\tau} \right) \quad \text{for }2x \ll
    \tau,
\end{equation}
and thus the mass accretion rate for the accretion ring $\dot m =
-2\pi Rv_{_R}\Sigma$ could be written as:
\begin{equation}
    \dot m \left(x,\tau \right) = \dot m_0
    \left( \frac{1}{2x}-\frac{2x}{\tau} \right) \frac{x^{3/4}}{\tau}\bm{e}^{-\frac{1+x^{2}}
    {\tau}}I_{1/4}\left( \frac{2x}{\tau} \right),
\end{equation}
 where $\dot m_0=6\pi\nu\Sigma_0$. This solution has the characteristic of "fast
rising" and ``slow falling'', which is similar to the observed X-ray
light curve. Fig 5 shows that it could fit well the observational
data. The fitting parameters: $\dot m_0=0.3$ M$_\odot$yr$^{-1}$,
$x=0.04$, $\tau_0=164$ days and $t_0$ is the 7th June of 2007.
$x=R/R_0=0.04$ indicates that the accretion rate obtained here
represents the mass accretion rate near the black hole. As shown
above, the simple clumpy accretion could produce the year-scale
variation of the X-ray luminosity observed in M87's nucleus. The
mass of gas clump $M_{\tiny\textrm{c}}\sim0.5$ M$_\odot$ could be
obtained through integration of $\dot m$ by time. According to
Wang et al. (2012), $R_0$ should be
$100R_{\tiny\textrm{Sch}}\sim1000R_{\tiny\textrm{Sch}}$.
  and typical density in gas clump ($n_{\tiny\textrm{cl}}$) is $\sim10^{14}/$cm$^{3}$. so we could estimate the radius of clump
$R_{\tiny\textrm{c}}=(3M_{\tiny\textrm{c}}/(4\pi
n_{\tiny\textrm{cl}}m_{\tiny\textrm{p}}))^{\frac{1}{3}}\sim
1.13\times10^{14}$ cm (here $m_{\tiny\textrm{p}}$ is the mass of
proton).

\begin{figure}
\center{\includegraphics[scale=0.6]{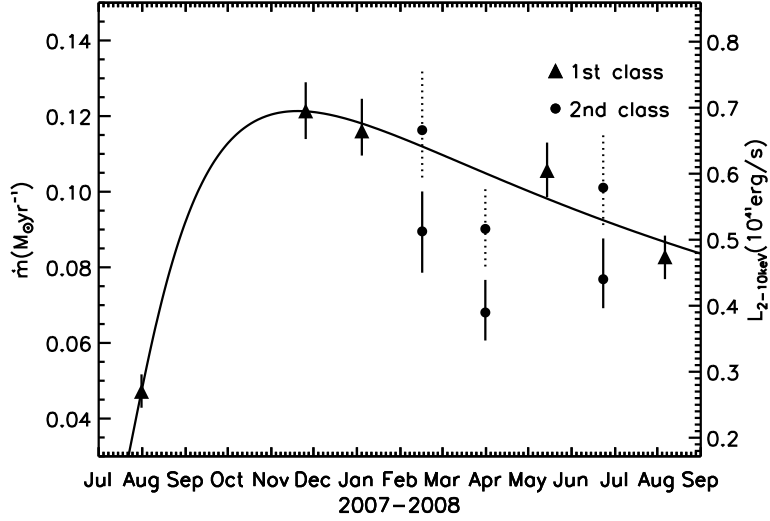}}
 \caption{The evolution of the mass accretion rate for M87's central
 blackhole. Triangles: the mass accretion rate/disk X-ray luminosity of the first class;
 Dots: the mass accretion rate/disk X-ray luminosity of the second class
 (The disk X-ray luminosity derived with $\Gamma_1=2.1$ is denoted with solid error bars and
  the disk X-ray luminosity derived with $\Gamma_1=2.0$ is denoted with dash error bars);
 Solid curve: the fitting result by clumpy accretion model.}\label{fig4}
\end{figure}

The anti-correlation of photon index and Eddington ratio in Fig 3
is also a prediction of the ADAF model (Yuan et al. 2007, Gu \&
Cao 2009). In ADAF, X-ray emission is mainly produced by the
comptonization of hot gas, as described in Gu \& Cao (2009), the
X-ray's photon index could be calculated by:
\begin{equation}\label{sigma_eq}
 \Gamma=-\ln~\tau_{\tiny\textrm{es}}/ln~A,
\end{equation}
here, $\tau_{\tiny\textrm{es}}$ is the electron scattering optical
length and could be expressed as
$\tau_{\tiny\textrm{es}}=24\alpha^{-1}\dot{m}r^{-\frac{1}{2}}$
($\alpha$ is the viscosity parameter $\sim0.1$, $\dot{m}$ is the
accretion rate in unit of Eddington rate, and the $r$ is the
radius in Schwarschild radii).
$A=16(k_{\tiny\textrm{B}}T_{\tiny\textrm{e}}/m_{\tiny\textrm{e}}c^2)^2$,
is the mean amplication factor of the photon energy by
Comptonization ($T_{\tiny\textrm{e}}$ and $m_{\tiny\textrm{e}}$ is
the temperature and mass of the hot electrons in ADAF disk)
(Rybicki \& Lightman 1979). So the photon index $\Gamma$ is
anti-correlated with $\dot{m}$. According to Equation 8, $\Gamma$
is determined by $\dot{m}$, $T_{\tiny\textrm{e}}$ and $r$.
$\dot{m}$ could be written as $\dot{\textrm{m}} =
L_{\tiny\textrm{x}}/(\eta c^2) $ and $\eta=10^{-5}$ (Di Matteo et
al. 2003). $r$ is taken as $50$R$_{\tiny\textrm{sch}}$ where the
accretion flow emits the detectable X-ray (Manmoto et al. 1997).
Then we adjust $T_{\tiny\textrm{e}}$ to produce observed $\Gamma$
and $\dot{m}$. We found that the $T_{\tiny\textrm{e}}$ is from
$2.5\times10^9$ K to $3.5\times10^9$ K. Such $T_{\tiny\textrm{e}}$
is consistent with the requirements of the ADAF model (Narayan \&
Yi 1994).

The light curve modelled by clumpy accretion is consistent well
with that observed in M87's nucleus. But such light curve could
also be explained by injection/acceleration of high energy
particles. Harris et al. (2003) suggests that the above process
could produce year-scale X-ray variations observed in HST-1.
Following this scheme, when the new energetic electrons is
injected or accelerated, the X-ray emission brightens up and the
spectra become harder. And then, the X-ray emission decreases by
radiative cooling or adiabatic expansion.

A way to differentiate the above mechanisms may be the variability
on multi-wavelength. For example, at the picture of particle
acceleration, the radio and X-ray emission will both brighten up
at the same time. Then, at the stage of synchrotron cooling, the
radio emission will persist longer than X-ray, because the cooling
rate is in proportion to $E^2$. However, for the picture of disk
evolution, the brightening up at long wavelength (e.g.
ultraviolet) should be in advance of that at short wavelength
(e.g. X-ray or hard X-ray). To acquire the light curves at
multi-wavelength in M87, a long-term joint monitoring from VHE to
radio band is being proposed (e.g., Harris et al. 2011 and Raue et
al. 2012).

\subsection{The X-ray components of flaring state: is the disk influenced by flare?}
Our results suggest that the X-ray emission of the second class
include both flaring and quiescent components. The spectra of
flaring components should be powerlaw-like with lower photon
index, for it is most likely emitted through the synchrotron
radiation of energetic electrons. In contrast, for quiescent
components, we could assume their spectra to be similar to the
first class. To decompose these two components, we fit the spectra
of the second class with an absorbed double-powerlaw as follows:
\begin{eqnarray}\label{eq2}\nonumber
Model_2=Wabs(n_{\tiny\textrm{H}})\times(Powerlaw(\Gamma_1,\textrm{norm}_1)\\
+Powerlaw(\Gamma_2,\textrm{norm}_2)),
\end{eqnarray}
the $\Gamma_1$ and norm$_1$ stands for the photon index and
normalization of the quiescent component; the $\Gamma_2$ and
norm$_2$ represent the photon index and normalization of the
flaring component.

We fix the $n_{\tiny\textrm{H}}$ as the $6.1\times10^{20}$
cm$^{-2}$. For the quiescent component, the observations of the
first class from 02/2008 to 08/2008 yield a photon index
($\Gamma_1$) $\sim-2.1$. The photon index of the flaring component
($\Gamma_2$) is hard to be constrained. Here, we consider the
coincidence between the VHE and X-ray flare in Feb 2008. The
photon index ($\alpha$) of VHE increment is about $-1.92$ (induced
from Albert et al. 2008) and the corresponding X-ray index is
$-1.46$ by $\Gamma=(\alpha-1)/2$ as a rough estimation.

In spectra fitting, the photon index is fixed as $\Gamma_1=2.1$
and $\Gamma_2=1.46$. The derived flux of quiescent component are
presented in Fig 5 by dots with solid error bars (Tab 2). Then we
also take the $\Gamma_1=2.0$ and $\Gamma_2=1.46$ and the derived
flux are shown as the dots with dotted error bars. Our result
reveals that the hard X-ray mainly comes from the flaring
component and the soft X-ray is dominated by quiescent component
(e.g. Fig 6). In three flaring states,the quiescent component only
accounts for the $25\%\sim40\%$ of the X-ray emission. The 2008
Feb.'s flare has the strongest flaring component almost twice as
bright as the others', in which a VHE flare was observed by MAGIC
and VERITAS (Albert et al. 2008, Acciari et al. 2009). For
outbursts occurring in Feb. and Jun. of 2008, the flux of
quiescent components is close to the prediction of year-scale
evolution($<2\sigma$) and it is lower in Apr. 2008.

\begin{table}
\begin{center}
\caption{The result of spectra fitting with absorbed double
powerlaw ($\Gamma_1=2.1$ and $\Gamma_2=1.46$). The error bars are
calculated with the confidence of $68\%$} \label{p125}
\begin{tabular}{@{}llllll}
\hline
 ObsID & Quiescent content & Flaring content& $\chi^2$/DOF ~~~\\
               & $10^{40}$ erg/s& $10^{40}$ erg/s&        \\
\hline
8577 &$5.1\pm0.6$  & $15.1\pm1.3$ & 64.99/65\\
8578 &$3.8\pm0.5$  & $7.5\pm1.0$  & 53.66/46\\
8580 &$4.4\pm0.6$  & $10.1\pm1.3$  &61.16/55\\
\hline
\end{tabular}
\end{center}
\medskip
\end{table}

\begin{figure}
\centering{\includegraphics[angle=-90,scale=0.4]{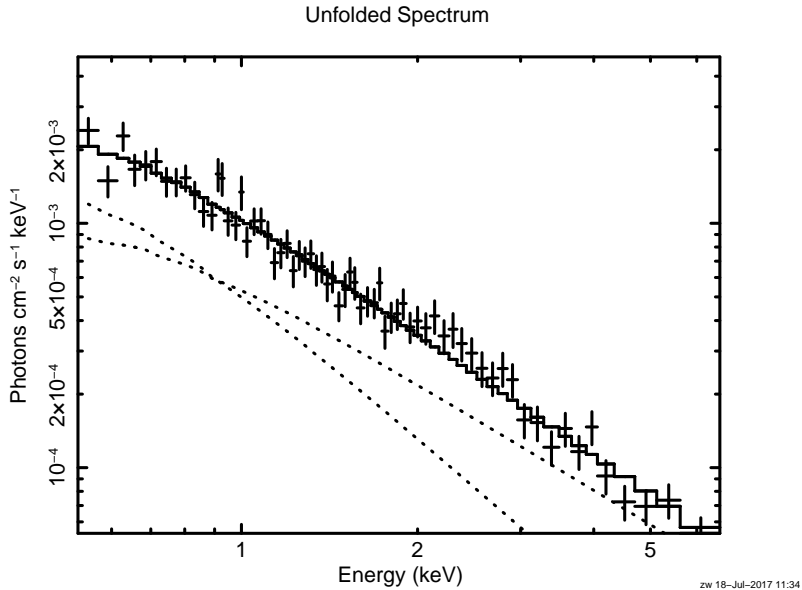}}
 \caption{The unfold spectra of M87's nucleus at flaring state(observation ID:8577, Reduced $\chi^2=0.9998$).}\label{fig2}
\end{figure}

To check the quiescent component in Apr 2008, in spectra fitting,
we fix $n_{\tiny\textrm{H}}=6.1\times10^{20}$ cm$^{-2}$,
$\Gamma_1=2.1$ and $\textrm{norm}_1=6.63\times10^{-4}$, which
means that the luminosity of quiescent component matches the curve
of the steady evolution. The resulting $\Gamma_2$,
$\textrm{norm}_2$ and reduced $\chi^2$ are 1.05, $1.47\times
10^{-4}$ and 1.38 respectively. The X-ray photon index of -1.05 is
too hard for the flaring component. Its corresponding index of the
electron's energy spectrum is about -1 ($\alpha=2\Gamma+1$), which
is far beyond the known accelerating mechanisms and difficult to
be explained by current models, such as Fermi acceleration
(Lieberman $\&$ Lichtenberg 1972) or stochastic acceleration (Fan
et al. 2010). Even in the case of $\Gamma_1=2.0$, the derived
luminosity in Apr. of 2008 is still lower than the others. So it
is possible that the flux of quiescent component in Apr. 2008
could not reach the value of year-scale evolution. The mechanism
for such a phenomenon is unclear. One possibility is that the disk
was disturbed by outburst if the flare occurred in the ADAF near
the black hole.

Our analysis reveals two kinds of variations for nuclear X-ray
emission: a year-scale variation and a month-scale variation. The
year-scale variation may come from the evolution of ADAF disk. And
the month-scale variation may be produced by the mini-jet. So the
process of jet-in-disk is favored for explaining the variations of
M87's nucleus. The mini-jet may be produced by the re-connection
of the magnetic field (Giannios et al. 2010). When magnetic
re-connection occurs, the local electrons in ADAF disk will be
accelerated to high energy. Correspondingly, X-ray spectra change
from the first class to the second class: the flaring component
brightens up and the quiescent component declines, as revealed in
$Chandra$ observation. Above process can only happen in the
magnetically arrested disk, which could be investigated in the
future by polarization detection.

\section{Conclusions}
\label{sect:conclusion}
In this paper, we re-analyze the M87's
nuclear spectra of 8 $Chandra$ observations from 2007 to 2008 and
discover a year-scale variation of the disk emission. Then we
discuss the evolution of accretion disk based on this year-scale
variation. At last, we also try to decompose the spectra of the
flaring states.

The conclusion are listed as follows:

1. We obtain both the month-scale and year-scale variations in
X-ray light curve of M87's nucleus. The month-scale variation is
produced by X-ray flares. The year-scale variation originates from
the evolution of accretion disk. Such a disk evolution could be
explained well by the accretion of a gas ring, which might result
from a tidal disrupted cloud of $\sim$0.5 M$_\odot$.

2. The X-ray spectra of M87's nucleus could be divided into two
classes: `quiescent state' and `flaring state'. The ADAF component
likely dominates the quiescent state. At flaring state, the X-ray
emission of nucleus includes both the ADAF component and `flaring
component', where the ADAF component accounts for $\sim$30$\%$.
But this fraction strongly depends on the assumed photon indexes.
More accurate observations are needed to constrain the results.

3. In flaring state, the ADAF component probably matches the value
of year-scale evolution. But in observation of Apr. 2008, the ADAF
component shows the trend of sharp decreasing, suggesting that
accretion flow was disturbed by the outburst occurring in the ADAF
of M87.
\section*{Acknowledgements}
 This work is supported by the CASSACA
Postdoc Grant (from the Chinese Academy of Sciences, CAS), the
Visiting Scholarship Grant (administered by the CAS South America
Center for Astronomy, CASSACA, NAOC), Science \& Technology Department of Yunnan Province - Yunnan University Joint Funding (2019FY003005)
and National Science Foundation of China (grant 11203019,11863006).

Cheng Cheng is supported by the Young Researcher Grant of National Astronomical Observatories,
 Chinese Academy of Science and the National Natural Science Foundation of China, No. 11803044.
 This work is sponsored (in part) by the Chinese Academy of Sciences (CAS),
through a grant to the CAS South America Center for Astronomy (CASSACA).



\label{lastpage}

\end{document}